\documentclass[useAMS]{mn2e}

\usepackage{amssymb,amsmath,psfig,times}
\usepackage{graphicx}

\def\gsim{ \lower .75ex \hbox{$\sim$} \llap{\raise .27ex \hbox{$>$}} }
\def\lsim{ \lower .75ex\hbox{$\sim$} \llap{\raise .27ex \hbox{$<$}} }

\def\sc{Schwarzschild}

\title[Precursors and GRB 091024]
{Afterglows from precursors in Gamma Ray Bursts. Application to the optical afterglow of GRB 091024}
\author[F. Nappo et al.]
{F. Nappo$^{1,2}$\thanks{Email: francesco.nappo@brera.inaf.it},
G. Ghisellini$^2$, G. Ghirlanda$^2$, A. Melandri$^2$, L. Nava$^3$, D. Burlon$^4$\\ \\
$^1$ Universit\`a degli Studi dell'Insubria, via Valleggio 11, I--22100 Como, Italy \\
$^2$ INAF -- Osservatorio Astronomico di Brera, Via Bianchi 46, I--23807 Merate, Italy\\
$^3$ Racah Institute for Physics, The Hebrew University of Jerusalem, 91904, Israel\\
$^4$ Sydney Institute for Astronomy, School of Physics, The University of Sydney, NSW 2006, Australia\\
}

\begin{document}  

\maketitle

\begin{abstract}
About 15\% of Gamma Ray Bursts have precursors, i.e. emission episodes preceding the main event,
whose spectral and temporal properties are similar to the main emission.
We propose that precursors have their own fireball, producing afterglow emission
due to the dissipation of the kinetic energy via external shock.
In the time lapse between the precursor and the main event, we assume that
the central engine is not completely turned off, but it continues to
eject relativistic material at a smaller rate, whose emission is below the background level.
The precursor fireball generates a first afterglow by the interaction with the external
circumburst medium. Matter injected by the central engine during the
"quasi-quiescent" phase replenishes the external medium with material in relativistic 
motion. The fireball corresponding to the main prompt emission episode 
rams into this
moving material,
producing a second afterglow, and finally catches up and merges with the first precursor fireball.
We test this scenario over
GRB 091024, an event with a precursor in the prompt light curve
and two well defined bumps in the optical afterglow, obtaining an excellent agreement with the existing data.
\end{abstract}

\section{Introduction}

Around 15\% of Gamma Ray Bursts (GRBs), show precursors 
in the light curve of their prompt emission.
Despite its definition is somewhat
subjective, a ``precursor"  can be identified as emission (i) preceding main event, 
with a maximum flux (or count rate) which is less than the main event, and (ii) separated from 
the latter by a time interval in which the flux is at the background level 
(for different definitions of precursors and different ways to find 
GRBs with precursors see Koshut et al. 1995, Lazzati 2005, Burlon et al. 2008, 2009).

The main question still to be addressed is whether precursors have a different origin than the main event in GRBs or not. 
This issue has been explored observationally by comparing the energetics and spectra of precursors with 
those of the main event and more in general with those of GRBs without precursors (e.g. Burlon et al. 2008, 2009). 
The spectrum and variability timescale of precursors are similar to those of the main event. 
On average, precursors have a fluence (flux integrated over its duration) about  one tenth of that of 
their associated main events. 
Also the time evolution of the spectrum of precursors is  consistent with that of GRBs without precursors 
(Burlon et al. 2009). 
These facts strongly suggest that the precursors are 
not a separated emission component, produced by something different
from the central engine responsible for the main event.
They are simply the first phases of the central engine activity.

The key question is then why the central engine
can have periods of time in which it is quiescent even for over 100 s in the rest frame of the source. 
Some GRBs showed more than one precursors, and precursors have been seen also in short 
bursts (a typical example is GRB 090510, a very bright 
short GRB observed also above 100 MeV by
the Large Area Telescope onboard the {\it Fermi} satellite, see
Ackermann et al. 2010; for other examples see Troja, Rosswog \& Gehrels 2010).

The explanation of the precursor phenomenon remains a puzzle.
The fact that there can be more than one precursor in one GRB rules out the 
``two step engine" model (see e.g. Wang \& M\'esz\'aros 2007; 
Lipunova et al. 2009) in which the precursor corresponds to the formation of a
compact object that after a while collapses into a black hole,
originating the main GRB event. The fact that the spectrum is 
not a blackbody (Lazzati et al. 2005; Burlon et al. 2008, 2009)
and that the quiescent times are often very long rule out the 
fireball precursor models (see e.g. M\'esz\'aros \& Rees 2000; 
Ruffini et al. 2001; Daigne \& Mochkovitch 2002; Lyutikov \& 
Blandford 2003; Li 2007) that associate the precursor 
emission to the thermal radiation initially trapped in the fireball.
Analogously, the non--thermal nature of the precursor's spectrum
rules out the ``progenitor precursor" models (Ramirez--Ruiz et al. 2002; 
Lazzati \& Begelman 2005) based on the collapsar scenario, 
in which the precursor corresponds to the interaction of 
a weakly relativistic jet with the stellar envelope. 

Recently, Bernardini et al. (2013, 2014) put forward a new model
able to explain the presence of one or more precursors with
non--thermal emission.
They assume that the central powerhouse is a newly born accreting magnetar;
in this model accretion can be halted by the centrifugal drag exerted by 
the rotating magnetosphere onto the infalling matter, allowing for multiple 
precursors and very long quiescent times. 

So far, however, precursors have been studied only in relation to the fact that 
they are observed in the high energy prompt light curve of GRBs. Here we consider
the possibility that some features of the afterglow emission can be ascribed to the 
precursor activity. In a nutshell, since the energy emitted by precursors is a large fraction
(10--50\%) of the total energy of the bursts, we can assume
that precursors originate fireballs as well as it is thought for the main event,
and that these fireballs run into the circumburst medium (CBM).

Therefore, precursors should produce an external shock, namely, an afterglow, similarly 
to what is thought to happen for the main emission episode of any GRB. 
When the quiescent time between the precursors and the main event
is sufficiently long,
as in the case of GRB 060124
(Romano et al.  2006) and GRB 050820A (Cenko et al. 2006), we have the opportunity to see
this afterglow emission (in between the precursor and the main event episodes).

The precursor fireball, being the first to interact with 
the CBM, will ``sweep" it and set it in motion with a certain velocity. 
Then the central engine enters a phase of quiescence,
ejecting matter at a much reduced rate, whose
emission remains below the background level.
This matter fills the circumburst region.
When the central engine becomes active again, producing the
main event of the prompt emission, a new fireball is produced,
running into the {\it moving} CBM ejected during the quiescence phase. 
The corresponding shock is weaker than the shock into a stationary
CBM, since the relative kinetic energy between the shock and the moving CBM is less
than in the case of an unperturbed and stationary CBM.
The main event fireball would then decelerate less, and it will  catch up with the
first fireball, shock it, and merge.

We apply these ideas to a specific burst, GRB 091024 (Gruber et al. 2011; Virgili et al. 2013), 
a very long burst ($T_{\rm 90}=1024$ s)  with a precursor lasting $\sim 70$ seconds, a quiescent time of
$\sim 600$ seconds, a second precursor lasting $\sim 40$ s, another 
quiescent time of $\sim 170$ seconds, and a complex main event, lasting
 $\sim 200$ seconds with pronounced variability.

The long duration and quiescent time allowed to have a dense
optical coverage of the light--curve, especially during the
quiescent time (Virgili et al. 2013).
Its properties make GRB 091024 an ideal test case for our new scenario, and
we anticipate that the results are very encouraging.

\begin{figure} 
\centering
\includegraphics[scale=0.5, angle=0]{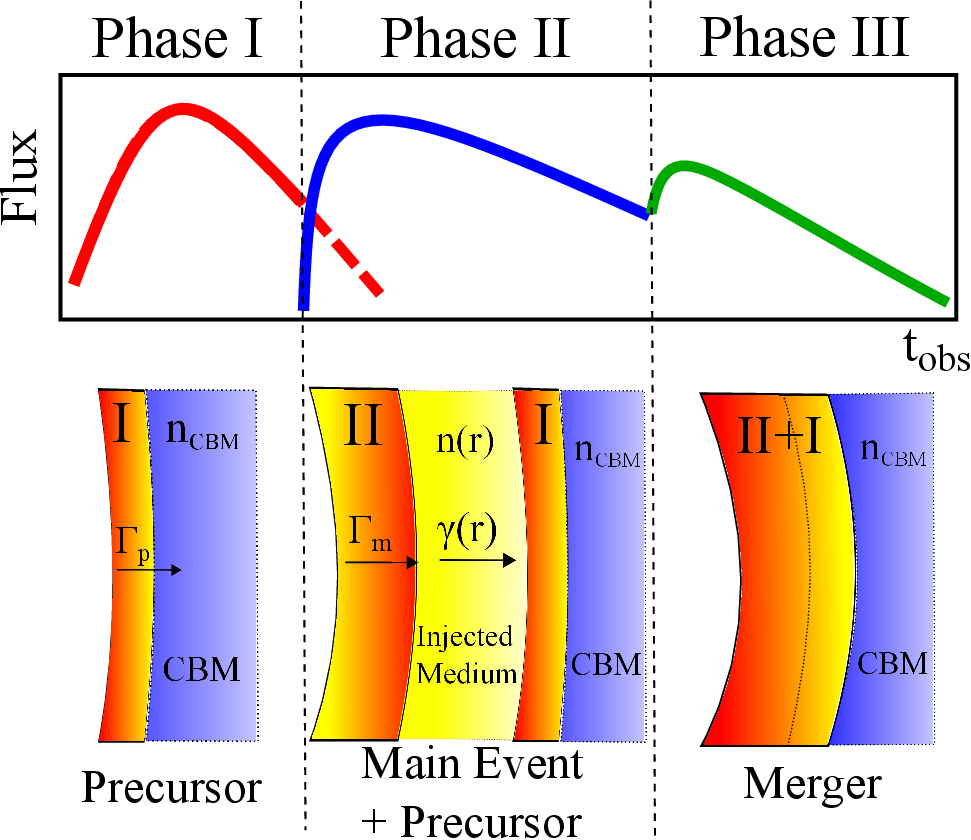}
\caption{Cartoon of the proposed scenario.
On the top we show the three phases of the
afterglow emission, that correspond to the
three configurations illustrated in the bottom:
Phase 1): Fireball I (associated to the precursor)
runs into the circumburst medium originating the first
afterglow.
Phase 2): Fireball II (associated to the main event)
runs into the moving material ejected by the central engine
during the quiescent phase, producing the second peak of the
afterglow light curve.
Phase III: Fireballs I and II merge, producing a brightening
of the afterglow.
}
\label{fig:cartoon}
\end{figure}

In \S2 we describe the main features of the proposed scenario, we detail its dynamical 
and radiative evolution in \S2.2, 2.3, 2.4. The model is then applied to the optical 
light curve of GRB 091024 in \S3 and \S4. We draw our conclusions in \S4. 

\section{Proposed scenario}

%
Burlon et al. (2008) showed that precursors have spectral and temporal 
properties similar to the main event emission, so it is reasonable to think that the
same central engine is responsible for both.
Fig. \ref{fig:cartoon} shows a cartoon of the proposed scenario,
whose main features can be summarized as follows:

\begin{itemize}

\item The precursor, being nothing else that the first pulse of the 
prompt emission, produces its own fireball (called ``fireball I" hereafter)  
that shocks the circumburst material (with density $n_{\rm CBM}$) producing an afterglow emission (the peak 
is shown by the solid red line in the top panel of Fig.\ref{fig:cartoon}).



\item During the ``quiescent" phase of apparent no emission, the central engine 
is not completely switched off, but it continues to eject 
relativistic matter, albeit at a much reduced rate. 
At the end of this quiescent phase, the central engine fully reactivates, producing 
the main prompt event associated to ``fireball II".

\item Fireball II can interact with this continuously ejected material and  produce a 
second afterglow emission. 
The shock will be less efficient than the first one because 
it is produced in a medium that is moving. 
This explains the presence of a second peak in the optical light curve, in this case associated 
with the second emitted fireball, i.e. that of the main event. 

\item For simplicity, the density and the bulk Lorentz factor profile of the 
ejected medium are assumed to be power--laws: 
\begin{equation}
\begin{matrix}
n(R) = n_0 (R/R_0)^{-s} \\
\gamma(R) = \gamma_0 (R/R_0)^{-g}
\end{matrix}
\label{eq:profili}
\end{equation}
\item If fireball I decelerates more than fireball II, by the interaction with the 
external medium, a merger between the two fireball occurs. 
During the merger some energy is released (with a mechanism that is similar to internal shocks) 
and a merged fireball (I+II) is formed. 
This fireball decelerates in the external medium generating the last part of the afterglow emission. 
This episode can be seen as an injection of new kinetic energy from the fireball II to the decelerating 
fireball I 
(\textit{refreshed shock}, Rees \& M\'esz\'aros 1998). 
This produces a rebrightening episode.

\end{itemize}
Summarizing, we identify three distinct afterglow emission phases:
\begin{enumerate}
\item The first afterglow produced by the deceleration of fireball I in the CBM.
\item The second afterglow produced by the deceleration of fireball II in the medium
ejected by the central engine, albeit at a reduced rate;
\item The last part of the afterglow produced by the merged fireballs (I+II) 
interaction with the external medium.
\end{enumerate}
For each of these three portions we need to study the dynamical behaviour of 
the fireball and how the dissipated energy is converted into radiation.

\subsection{Dynamics}

This work requires a non--standard dynamical treatment because the afterglow is produced by the interaction of a relativistic fireball with a medium that is \textit{moving} with respect to the central engine. We propose a new approach, inspired by Nava et al. (2013), based on the energy--momentum conservation in the collision of the fireball with the external CBM, that can be in motion or at rest.
Here we consider the case of an external medium that is moving with a Lorentz factor profile $\gamma(R)$.
We assume that the fireball has initial mass $M_0$ and initial Lorentz factor $\Gamma_0$.
In the collision with the external CBM,
the fireball collects the mass of the encountered CBM and when the 
collected mass is large enough, the fireball starts to decelerate, reducing its 
bulk Lorentz factor $\Gamma(R)$.

In the collision, some bulk kinetic energy is transformed into internal energy $\varepsilon'$ 
(as measured in the fireball comoving frame). 
A fraction of this internal energy (labeled as $\varepsilon^\prime_{\rm RAD}$) is radiated 
and no longer contributes to the inertia of the fireball; the amount of radiated energy depends on 
the specific radiative process and it will be analysed in the following section. 

If $\varepsilon^\prime_{\rm RAD} \ll \varepsilon^\prime$, the fireball is a very inefficient radiator, 
so the radiative losses are negligible and the dynamical evolution of the fireball can 
be described separately by the radiative emission behaviour. 
Since the total fireball energy is conserved, this case is called \emph{adiabatic regime}.

In the opposite case, if $\varepsilon^\prime_{\rm RAD} \sim \varepsilon^\prime$, 
almost all the internal energy is lost in radiation. 
Also in this case the fireball dynamical evolution can be treated separately by the radiative 
emission, and it is called \emph{radiative regime}.

In the intermediate case, the dynamical evolution is strictly linked to the radiative behaviour, 
so both aspects have to be considered simultaneously.

%

To describe the dynamical evolution of the fireball and to compute the amount of dissipated energy we solved numerically the equations of the energy--momentum conservation step--by--step. 
At radius $R$, the fireball with Lorentz bulk factor $\Gamma(R)$ and mass $M(R)$ hits an element of external medium $\Delta m=4\pi R^2 n(R) m_p \Delta R$ that is moving with a Lorentz factor $\gamma(R)$; from energy--momentum conservation, we can obtain the value of the dissipated energy in the collision $\Delta \varepsilon^\prime (R)$ and the new Lorentz factor $\Gamma (R + \Delta R)$.

A fraction $\epsilon_e$ of the dissipated energy is given to the CBM electrons, that can emit 
a part of it as radiation. We assume that we can write the 
radiated energy $\Delta \varepsilon'_{\rm RAD}$ as:
\begin{equation}
\label{eq:epsRAD}
\Delta \varepsilon^\prime_{\rm RAD}(R) = \zeta(R) \epsilon_e \Delta \varepsilon^\prime(R)
\end{equation}
where $\zeta(R)$ is the fraction of electron energy that is radiated at $R$ 
(this value depends on the radiative process and it will be calculated in \S2.3).
We obtain the comoving emitted bolometric luminosity:
\begin{equation}
L^\prime_{\rm bol} = \frac{\Delta \varepsilon^\prime_{\rm RAD}}{\Delta t^\prime}=
\frac{\Delta \varepsilon^\prime_{\rm RAD}}{\Delta R}
\frac{\Delta R}{\Delta t_{\perp}}\frac{\Delta t_{\perp}}{\Delta t'}=
\Gamma \beta c \frac{\Delta \varepsilon^\prime_{\rm RAD}}{\Delta R}
\end{equation}
where $\Delta t_{\perp}=\Delta R /(\beta c)$ is the time interval corresponding to the 
distance $\Delta R$ measured by an observer placed perpendicularly to the burst direction, 
$\Delta t'$ is the same time measured by a comoving observer,
$\beta$ is the bulk velocity of the fireball (in unit of $c$) and $\Gamma$ is the bulk Lorentz factor of the fireball.

\subsection{Times}

The time interval $\Delta t_{\rm obs}$ corresponding to 
the distance $\Delta R$ measured by an observer placed close to the axis of the burst can be written as:
\begin{equation}
\Delta t_{\rm obs} =(1+z) \frac{1-\beta \cos \theta}{\beta c}\Delta R
\label{eq:deltat0}	
\end{equation}
where $z$ is the cosmological redshift and $\theta$ is the angle between 
the line of sight and the burst axis. 

The observed bolometric luminosity observed at time $t_{\rm obs}$ is constituted by the photons  
arriving to the detector at the time $t_{\rm obs}$, but that have been emitted
at different times.
The real emitting surface seen at time $t_{\rm obs}$ is called \emph{equal arrival 
time surface} and it is non--trivial to determine. 

We approximate the determination of the equal arrival time surfaces by
considering that most of the received emission comes from the ring of aperture angle 
$\sin \theta = 1/\Gamma$ (or equivalently $\cos \theta = \beta$). 
In this way Eq. \ref{eq:deltat0} simplifies in
%
\begin{equation}
\Delta t_{\rm obs} = (1+z)\frac{\Delta R}{\beta c\Gamma^2}
\label{eq:deltat1}
\end{equation}
The observed bolometric luminosity can be written as:
\begin{equation}
L_{\rm bol} = \delta^2 L^\prime_{\rm bol}
\end{equation}
where $\delta = \left[\Gamma(1-\beta \cos \theta)\right]^{-1}$ 
(equal to $\Gamma$ for $\cos \theta = \beta$) is the relativistic Doppler 
factor (or beaming factor).
The obtained bolometric luminosity is a function of time, so we will label it with $L_{\rm bol}(t)$.

\subsection{Radiative emission}

In this section we calculate the synchrotron emission observed at the
frequency $\nu= \nu_{\rm obs}(1+z)$.
We propose a new method based on the normalization of the synchrotron spectrum to the bolometric luminosity computed in the previous sections. Most of the equations used in this section are inspired by the work of Panaitescu \& Kumar (2000),  generalizing them in order to describe the emission produced by the interaction with a medium that has been pre--accelerated to relativistic speed.

We explicitly apply an energy conservation equation,
by requiring that in the fast cooling regime the frequency integrated spectrum
is equal to the power that the shock gives to the electrons, as measured in the
observer frame:
\begin{equation}
\int L_{\nu}(t)\mbox{d}\nu = \Gamma^2
\epsilon_{\rm e} {\Delta\varepsilon^\prime \over \Delta t^\prime}
\end{equation}
We assume that the synchrotron emission is produced by a distribution of electrons
that is the result of a continuous particle injection and the cooling due to the emission. 
We label with $Q(\gamma)$ the number of injected electrons with energy 
$\gamma m_e c^2$ for unity of time and volume, $N(\gamma)$ the energetic distribution 
for unity of volume and $\dot{\gamma}$ the cooling rate. 
They are linked by  the \textit{continuity equation:} 
\begin{equation}
\label{eq:cont}
\frac{\partial N(\gamma)}{\partial t}=
\frac{\partial}{\partial \gamma}\left[ \dot{\gamma} N(\gamma)\right] + Q(\gamma)
\end{equation}
For a stationary source $\partial N\left(\gamma\right)/\partial t = 0$.\\
We assume that the injection distribution is a power law of index $2<p<3$:
\begin{equation}
\label{eq:powerlaw}
Q(\gamma) \propto \gamma^{-p} \mbox{ for } \gamma_i < \gamma < \gamma_{\rm max},
\end{equation}
where $\gamma_i$ is called injection energy.

The synchrotron cooling rate is:
\begin{equation}
\label{eq:syncool}
\dot{\gamma}_{\rm syn} m_{\rm e} c^2=\frac{4}{3}\sigma_{\rm T} c U_B \gamma^2
\end{equation}
The synchrotron cooling time is (Sari et al., 1998):
\begin{equation}
\label{eq:cooltime}
t_{\rm c,syn} = \frac{\gamma}{\dot{\gamma}_{syn}}=\frac{3m_{\rm e} c}{4\sigma_{\rm T} U_B \gamma}
\end{equation}
For each time, we define the cooling energy $\gamma_c(t) m_{\rm e} c^2$, that is the energy that an 
electron must have so that its cooling time is equal to the scale time of expansion of the fireball.
Under these conditions we have two possible solutions of Eq. \ref{eq:cont} (see also Sari \& Esin, 2001):

\begin{itemize}

\item[a)] \textit{Fast Cooling regime}: $\gamma_{\rm c} < \gamma_{\rm i}$. 
In this case all injected electrons cool in a time shorter than the
dynamical time (time needed to double the fireball radius).
The corresponding energy distribution $N_{\rm FC}(\gamma)$ is  

\begin{equation}
\label{eq:NFC}
N_{\rm FC}(\gamma) =
\begin{cases}
 0  & \gamma<\gamma_{\rm c}, \\
 N_{\rm i} \left(\frac{\gamma}{\gamma_{\rm i}}\right)^{-2} & \gamma_{\rm c} < \gamma < \gamma_{\rm i}, \\
 N_{\rm i} \left(\frac{\gamma}{\gamma_{\rm i}}\right)^{-(p+1)} & \gamma_{\rm i} < \gamma < \gamma_{\rm max} \\
\end{cases} 
\end{equation}

\item[b)] \textit{Slow Cooling regime}: $\gamma_{\rm i} < \gamma_{\rm c}$. 
In this case a fraction of (or all) the injected electrons do not cool in a dynamical time.
The corresponding energy distribution $N_{\rm SC}(\gamma)$ is  
\begin{equation}
\label{eq:NSC}
N_{\rm SC}(\gamma) =
\begin{cases}
 0  & \gamma<\gamma_{\rm i}, \\
 N_{\rm c} \left(\frac{\gamma}{\gamma_{\rm c}}\right)^{-p} & \gamma_{\rm i} < \gamma < \gamma_{\rm c}, \\
 N_{\rm c} \left(\frac{\gamma}{\gamma_{\rm c}}\right)^{-(p+1)} & \gamma_{\rm c} < \gamma < \gamma_{\rm max} \\
\end{cases} 
\end{equation}

In this work we will not derive the normalizations $N_{\rm i}$ and $N_{\rm c}$ of the electrons energy distribution, but we will compute directly the normalizations of the emitted synchrotron spectrum (see \S \ref{sec:spec}).

\end{itemize}

\subsubsection{Break frequencies}\label{sec:break}

Generalizing Panaitescu \& Kumar (2000) in the case of a pre--accelerated 
medium\footnote{Note the different use of $\Gamma$ and 
$\Gamma_{\rm rel}$.
Here $\Gamma$ is the bulk Lorentz factor of the fireball measured by the observer, 
while $\Gamma_{\rm rel}$ is the bulk Lorentz factor of the fireball as measured in
the frame of the injected moving medium.
}
we have the following relations:
\begin{equation}
\label{eq:gammai}
\gamma_{\rm i}  =  \epsilon_e\, { m_{\rm p}\over m_{\rm e} }\, {p-2 \over  p-1}\,
(\Gamma_{\rm rel}-1) 
\end{equation}
\begin{equation}
\label{eq:gammac}
\gamma_c=\frac{15\pi}{1+y} \, \frac{m_{\rm e} c^2\Gamma}{\sigma_{\rm T} B^2 R}
\end{equation}
\begin{equation}
\label{eq:campomag}
U_B=\frac{B^2}{8\pi}=\epsilon_B n(R) m_p c^2(\Gamma_{\rm rel}-1)(4\Gamma_{\rm rel}+3)
\end{equation}
\begin{equation}
\label{eq:yPK}
y=\frac{4}{3}\tau_{\rm e} \left\langle \gamma^2 \right\rangle
\end{equation}
\begin{equation}
\label{eq:taue}
\tau_e = \frac{1}{4\pi}\frac{\sigma_{\rm T} m(r)}{m_{\rm  p} R^2}
\end{equation}
where $U_B$ is the magnetic energy density, $y$ is the Comptonization parameter, $\tau_{\rm e}$ is 
the electron optical depth, $\left\langle \gamma^2 \right\rangle$ is the average of the square of the random electron Lorentz factor, 
$\Gamma_{\rm rel}$ is the relative Lorentz factor between the fireball and the external 
medium, as measured in one of the two frames, 
and $\epsilon_{\rm  e}$, $\epsilon_B$ are the fractions of the dissipated internal energy 
that are given respectively to leptons and to the magnetic field.
The injection and the cooling frequency are:
\begin{equation}
\label{eq:nui}
\nu_{\rm i}=\frac{4}{3}\frac{e B}{2\pi m_{\rm e} c}\Gamma\gamma_{\rm i}^2
\end{equation}
\begin{equation}
\label{eq:nuc}
\nu_{\rm c}=\frac{4}{3}\frac{eB}{2\pi m_{\rm  e} c}\Gamma\gamma_c^2
\end{equation}
For simplicity, we will not consider the self--absorbed part of the spectrum 
when calculating the normalization, 
since it is less relevant for the total energy. 
Anyway we will compute the self--absorption frequency $\nu_{\rm a}$ using Eq. 52 
in Panaitescu \& Kumar (2000).
%
The equation used is:
\begin{equation}
\label{eq:selfabs}
\nu_{\rm a} = \nu_{\rm m}\left(\frac{3-s}{5}\frac{B \gamma_{\rm m}^5}{e n R}\right)^{-\alpha} \mbox{,} \quad
\alpha =
\begin{cases}
-\frac{3}{5} & \nu_{\rm a} < \nu_{\rm m} \\
-\frac{2}{q+4} & \nu_{\rm a} > \nu_{\rm m} \\
\end{cases}
\end{equation}
where $\nu_{\rm m} = \min(\nu_{\rm i}, \nu_{\rm c})$, $\gamma_{\rm m} = \min(\gamma_{\rm i}, 
\gamma_{\rm c})$ and $q=2$ in \textit{fast cooling} regime $(\nu_{\rm c} < \nu_{\rm i})$ 
or $q=p$ in \textit{slow cooling} regime $(\nu_{\rm i} < \nu_{\rm c})$. 

The fraction of emitted energy $\zeta$ (cfr. Eq. \ref{eq:epsRAD}) can be expressed as the 
ratio between the power emitted in slow cooling regime and in fast cooling regime. 
\begin{equation}
\label{eq:zeta}
\zeta = \frac{\left[\int N(\gamma)\dot{\gamma}\mbox{d}\gamma\right]_{\rm SC}}
{\left[\int N(\gamma)\dot{\gamma}\mbox{d}\gamma\right]_{\rm FC}}=
\begin{cases}
\frac{\gamma_{\rm i}}{\gamma_{\rm c}}\frac{p-2}{3-p}\left[\frac{1}{p-2}
\left(\frac{\gamma_c}{\gamma_{\rm i}}\right)^{3-p}-1\right] & \gamma_{\rm i} \leq \gamma_{\rm c}, \\
1 & \gamma_{\rm c}  < \gamma_{\rm i} \\
\end{cases} 
\end{equation}
where we used $\dot{\gamma}=\dot{\gamma}_{\rm syn}$.
Obviously, in the fast cooling regime we must have $\zeta = 1$.

\subsubsection{Comptonization parameter $y$}

We introduce here a new way to estimate the Comptonizazion parameter $y$.
The bolometric luminosity is constituted by two main components: 
the synchrotron emission and the SSC.
\begin{equation}
\label{eq:lum}
L^\prime_{\rm syn} + L^\prime_{\rm SSC} = L^\prime_{\rm bol}  = \zeta(y) P^\prime_{\rm e} 
\end{equation}
where $P^\prime_{\rm e}$ is the total power distributed to the electrons, i.e. 
(cfr. Eq. \ref{eq:epsRAD}): 
\begin{equation}
P^\prime_{\rm e} = \epsilon_{\rm e} \frac{\Delta\varepsilon^\prime}{\Delta t^\prime}
\end{equation}
$P^\prime_{\rm e}$ is obtained from purely dynamical considerations and does not depend on 
the particular emission mechanism chosen.
We can express the synchrotron luminosity as (see Ghisellini et al. 2010):
\begin{equation}
L^\prime_{\rm syn} = \int 4 \pi R^2 \Delta R^\prime  N(\gamma) \dot{\gamma}_{\rm syn}m_{\rm e} c^2 \mbox{d}\gamma
\label{eq:GG1}
\end{equation}
Following Eq. \ref{eq:yPK} we have
\begin{equation}
y =  {4\over 3}\sigma_{\rm T} \Delta R^\prime \int N(\gamma)\gamma^2 d\gamma
\end{equation}
The we have:
\begin{equation}
\label{eq:lumsyn}
L^\prime_{\rm syn} = 4\pi R^2 c U_B y
\end{equation}
Dividing both sides for $4\pi R^2 c$ we obtain the energy density of the synchrotron emission:
\begin{equation}
\label{eq:ensyn}
U_{\rm syn} = \frac{L^\prime_{\rm syn} }{4\pi R^2 c}= y U_B 
\end{equation}
Adopting again Eq. \ref{eq:GG1} with 
$\dot{\gamma}_{\rm SSC} = (4/3)\sigma_{\rm T} c \gamma^2 U_{\rm syn}/(m_{\rm e} c^2)$ we can express the SSC luminosity as:
\begin{equation}
\label{eq:lumssc}
L^\prime_{\rm SSC}  = 4\pi R^2 c U_{\rm syn} y = y L^\prime_{\rm syn}  = y^2 4\pi R^2 c U_B
\end{equation}
valid as long as the scattering is in the Thomson regime.
Combining the Eqs. \ref{eq:lum}, \ref{eq:lumsyn} and \ref{eq:lumssc} we
obtain an important relation:
\begin{equation}
\label{eq:ips}
4\pi R^2 c U_B (y+y^2) = \zeta (y) P_e'
\end{equation}
We obtain the Comptonization parameter $y$ by numerically solving:
\begin{equation}
\label{eq:ypsilon}
\frac{y+y^2}{\zeta (y)} = \frac{P^\prime_{\rm e}}{4\pi R^2 c U_B}
\end{equation}

\subsubsection{Normalized synchrotron spectrum}\label{sec:spec}

The synchrotron spectrum can be written as Eq. \ref{eq:speFC} or Eq. \ref{eq:speSC}, 
in the fast and slow cooling regime respectively (see also Sari et al, 1998, for the shape of the spectrum).
\begin{equation}
\label{eq:speFC}
L_{\rm \nu, syn}^{FC} = A_{\rm FC}
\begin{cases}
 \left( \frac{\nu}{\nu_{\rm c}} \right)^{1/3} & \nu < \nu_{\rm c} \\
 \left( \frac{\nu}{\nu_{\rm c}} \right)^{-1/2} & \nu_{\rm c} < \nu < \nu_{\rm i} \\
 \left( \frac{\nu}{\nu_{\rm i}} \right)^{-p/2}\left( \frac{\nu_{\rm c}}{\nu_{\rm i}} \right)^{1/2} 
 & \nu_{\rm i} < \nu < \nu_{\rm max} \\
\end{cases}
\end{equation}
\begin{equation}
\label{eq:speSC}
L_{\rm \nu, syn}^{\rm SC} = A_{\rm SC}
\begin{cases}
 \left( \frac{\nu}{\nu_{\rm i}} \right)^{1/3} & \nu < \nu_{\rm i} \\
 \left( \frac{\nu}{\nu_{\rm i}} \right)^{-(p-1)/2} & \nu_{\rm i} < \nu < \nu_{\rm c} \\
 \left( \frac{\nu}{\nu_{\rm c}} \right)^{-p/2}\left( \frac{\nu_{\rm i}}{\nu_{\rm c}} \right)^{(p-1)/2} 
 & \nu_{\rm c} < \nu < \nu_{\rm max} \\
\end{cases} 
\end{equation}
where $A_{\rm FC}$ and $A_{\rm SC}$ are normalizing constants. 
In Eqs. \ref{eq:speFC} and \ref{eq:speSC}, 
we neglected the self-absorbed part of spectrum, because we do not consider the evolution of 
the radio afterglow. 
We can obtain the normalizations imposing:
\begin{equation}
L_{\rm syn}+L_{\rm SSC} =L_{\rm syn}(1+y)=L_{\rm bol}
\end{equation}
or
\begin{equation}
\label{eq:normal}
\int L_{\rm \nu,syn} \mbox{d} \nu = \frac{L_{\rm bol}}{1+y}
= \zeta {\Gamma^2 \epsilon_{\rm e} \over 1+y} {\Delta\epsilon^\prime \over \Delta t^\prime}
\end{equation}
where $L_{\rm bol}$ is obtained from the dynamical energy--momentum conservation.
Remembering that $\nu=\nu_{\rm obs} (1+z)$, the observed 
monochromatic flux can be written as:
\begin{equation}
\label{eq:fluxmono}
F_{\nu_{\rm obs}} = \frac{\nu}{\nu_{\rm obs}}\frac{L_{\nu}}{4\pi d_L^2}=(1+z)\frac{L_{\nu}}{4\pi d_L^2}
\end{equation}
where $d_L$ is the luminosity distance.
Determining for each time the specific flux $F_{\nu_{\rm obs}}$, we obtain the monochromatic 
flux light curve, i.e. the function $F_{\nu_{\rm obs}}(t)$.

With the above approach we 
obtain the monochromatic synchrotron light curves 
of the GRB afterglow in a way that ensures the conservation of energy and momentum. 

\subsection{Model parameters and observables}\label{sec:param}

We can divide the model parameters in the following six groups.
\vskip 0.2 cm
\noindent
\textbf{Precursor parameters}
\begin{itemize}
\item \textbf{$\Gamma_{\rm 0,p}$}: Initial Lorentz factor of the precursor fireball
\item \textbf{$E_{\rm k,p}$}: Isotropic equivalent kinetic energy of the precursor fireball
\item \textbf{$T_{\rm p}$}: Duration of the precursor emission
\item \textbf{$E_{\rm peak, p}$}: Peak energy of the precursor 
\end{itemize}
The last two quantities are measurable and the kinetic energy $E_{\rm k,p}$ can be obtained 
by the isotropic equivalent energy of the precursor $E_{\rm ISO,p}$ from the relation:
$E_{\rm ISO,p}={\eta} E_{\rm k,p}$,
where $\eta$ is the efficiency of conversion of the kinetic energy in radiative energy.
We will assume $\eta \simeq 0.2$ (Frail et al., 2001 and see Ref. in Ghirlanda et al., 2004). 
The isotropic equivalent energy can be obtained from 
the observed fluence $\mathcal{F}$ with:
\begin{equation}
E_{\rm ISO} = {4\pi d_L^2 \over (1+z)}\mathcal{F}
\end{equation}
So the only free parameter of the precursor is the initial Lorentz factor $\Gamma_{\rm 0,p}$.

\vskip 0.2 cm
\noindent
\textbf{Circumburst medium density parameters}

The density profile of the external circumburst medium strongly influences the shape of the afterglow light curve.
In the general case, the density profile can be written as:
$$n(R) = A R^{-s}$$
In the literature, two cases are usually considered: $s=0$ (homogeneous profile), $s=2$ (wind profile).

Following Sari (1997) and Ghisellini et al. (2010),  the observed bolometric luminosity is:
$$L \propto \Gamma^8 t^2 n$$
with $R \propto t \Gamma^2$.

In the \textit{coasting phase} $\Gamma$ is constant, implying $L \propto t^2 n$ and $R \propto t$.
If the medium is homogeneous, we then have $L \propto t^2$.
Instead, for a wind medium, $L \propto t^0$ since the increase of the observable surface ($\propto R^2 \propto t^2$)
is compensated by the decreasing density ($n \propto R^{-2}\propto t^{-2}$).

The model is valid with both density profiles.
When an initial rise of the early afterglow optical flux is observed, we will assume 
(in the following application of the model) that the precursor propagates in a homogeneous circumburst medium.
In this case the density of the circumburst medium $n$ is another free parameter of the model.

\vskip 0.2 cm
\noindent
\textbf{Main event parameters}
\begin{itemize}
\item \textbf{$\Gamma_{\rm 0,m}$}: Initial Lorentz factor of the main event fireball 
\item \textbf{$E_{\rm k,m}$}: Isotropic equivalent kinetic energy of the main event fireball
\item \textbf{$T_{\rm m}$}: Duration of the main event emission
\item \textbf{$E_{\rm peak, m}$}: Peak energy of the spectrum of the main event
\item \textbf{$\Delta t_{\rm main}$}: Start time of the main event emission after the trigger
\end{itemize}
$\Delta t_{\rm main}$ can be directly observed. As in the previous case, the only 
free parameter is the initial Lorentz factor $\Gamma_{\rm 0,m}$.

\vskip 0.2 cm
\noindent
\textbf{Injected medium parameters}

The  matter injected by the central engine in the quasi--quiescence phases can be described
with 4 parameters characterizing the density $n(R)$ and velocity $\gamma(R)$ profile, 
i.e. $\gamma_0$, $g$, $n_0$ and $s$ (see Eq. \ref{eq:profili}).

\vskip 0.2 cm
\noindent
\textbf{Merging parameters}

These parameters describe the third phase of the afterglow light curve, after the merging 
of fireball I and II.
\begin{itemize}
\item \textbf{$t_{\rm merg}$}: Merging time of the two fireballs
\item \textbf{$\Gamma_{\rm merg}$}: Initial Lorentz factor of the merged fireball
\item \textbf{$E_{\rm k,merg}$}: Isotropic equivalent kinetic energy of the merged fireball
\end{itemize}
In this group there are no free parameters. 
The merging time $t_{\rm merg}$ can be determined
studying the motion of the fireballs I and II that is completely determined from the 
parameters $\Gamma_{\rm p}(t)$, $\Gamma_{\rm m}(t)$, and the medium properties.
$\Gamma_{\rm merg}$ and $E_{\rm k,merg}$ can be obtained from the equation of energy--momentum 
conservation during the merging.

\vskip 0.2cm
\noindent
\textbf{Energy distribution parameters}

For each part of the afterglow light curve, there are three free parameters:
\begin{itemize}
\item \textbf{$\epsilon_{\rm e}$}: Fraction of the internal energy given to  electrons
\item \textbf{$\epsilon_{\rm B}$}: Fraction of the internal energy given to  magnetic field
\item \textbf{$p$}: Injection spectral index of the electrons 
\end{itemize}
Obviously we must have $\epsilon_{\rm e} + \epsilon_{\rm B} < 1$.

\vskip 0.2 cm
The model has many free parameters (16, considering also the density $n$ of the circumburst medium),
but they can be 
determined or limited by the observational data.
The position of the peaks and the slopes in the light curves are the main observables that can 
constrain the parameters value. 
In particular we have:
$(t_{\rm p}, F_{\rm \nu, p})$ that are respectively the time and flux of the peak of the precursor afterglow, 
$(t_{\rm m}, F_{\rm \nu, m})$ of the peak of the main event afterglow, $(t_{\rm merg}, F_{\rm \nu, merg})$ that 
are the time and flux corresponding to the moment of the merger between fireball I and II.

From the slopes (especially for the second afterglow peak) we can infer the properties of the medium
injected by the central engine (see also the application in \S \ref{sec:optres} and Fig. \ref{fig:par}). 

\subsection{Observable afterglow peaks and timescales}\label{sec:timi}
The three--peak afterglow scenario that we described should be the more common case, but other
outcomes are possible.
In fact, varying the time separation between precursor and main event fireballs, or modifying the properties of 
the injected medium we can reproduce different behaviours. 

In the case at hands, the first peak in the afterglow light curve is interpreted as due to the deceleration of 
the precursor fireball in the external medium. The time of the onset of the precursor afterglow is strictly 
correlated with the deceleration time $t_{\rm dec} \propto (E_{\rm k}/n \Gamma_0^8)^{1/3}$ (this relation is 
valid only for a homogeneous CBM; see Ghirlanda et al., 2012 and Nava et al., 2013 for more details). 
This requires that the fireball launched by the main event does not catch up the fireball of the
precursor before the latter has decelerated.
This can never happen if $t_{\rm dec}$ of the precursor fireball is shorter than the precursor--main event
time interval.
Instead, if the two fireballs interact before the deceleration time, the first afterglow peak 
disappears.

Knowing the density of the interstellar medium and the kinetic energy of the fireball, we can easily link 
the deceleration time to the initial bulk Lorentz factor. 
The presence of an optical peak between the precursor and the main event 
prompt emissions is key to constrain the value of the initial bulk Lorentz factor of the precursor. 
If the peak is not observed, 
we have $t_{\rm dec}(\Gamma_{0, \rm p}) > \Delta t_{\rm main}$.

We cannot simply derive the deceleration time of the second fireball, since 
the main event propagates in a medium that is moving. 
This implies that it is difficult to constrain the initial bulk Lorentz factor of the main event, 
since its estimate depends on the properties of the injected medium.

The second peak is suppressed if the second fireball does not decelerate before
reaching the first fireball.
This occurs when the injected medium is moving too fast or has very low density. 

\section{Application to GRB 091024}

As an application of our model,
we study the broad band emission of GRB 091024, an extremely long GRB,
with $T_{90} \sim 1,020$ s with precursor and a double peaked optical light curve. 
This burst is particularly suited to test our proposed model also because its optical 
light curve is highly sampled from 100 s to 1 day after the prompt emission.
It was first  detected by Fermi/GBM, that triggered at 08:55:58.47 UT ($t_0$) and again 
at 09:06:29.36 UT (Bissaldi \& Connaughton, 2009).
It was observed also by Konus--Wind (Golenetskii et al. 2009), and by 
{\it Swift}/BAT (Marshall et al., 2009),
but the burst exited the BAT field of view at $t \sim 460$ s after the 
trigger and XRT measurements are available only $\sim 3,000$ s after the GBM trigger,
because of Earth--limb constraints. 
The burst coordinates, according to XRT, are $\alpha_{J2000}=22^h37^m00.4$, 
$\delta_{J2000}=56^\circ 53' 21''$, with an accuracy of 6 arcsec (Page \& Marshall, 2009).
{\it Fermi}/LAT was re--pointed at the burst direction, but it 
did not reveal any significant emission (Bouvier et al., 2009).

The acquisition of the first optical data started about a minute after 
the first trigger by the Super--LOTIS telescope (Updike et al., 2009).
Other photometric data come from the Sonoita Research Observatory (SRO) and the 
Faulkes telescopes.
The afterglow optical spectrum was obtained with the Low Resolution Imaging Spectrometer 
on telescope Keck I and by the spectrograph GMOS-N on Gemini North, measuring a 
cosmological redshift of $z= 1.092$ (Cenko et al., 2009; Cucchiara et al., 2009).

The optical data that we use for the physical interpretation 
of the afterglow are published in Virgili et al. (2013).

\subsection{Prompt emission}

\begin{figure} 
\centering
\includegraphics[scale=0.3, angle=0]{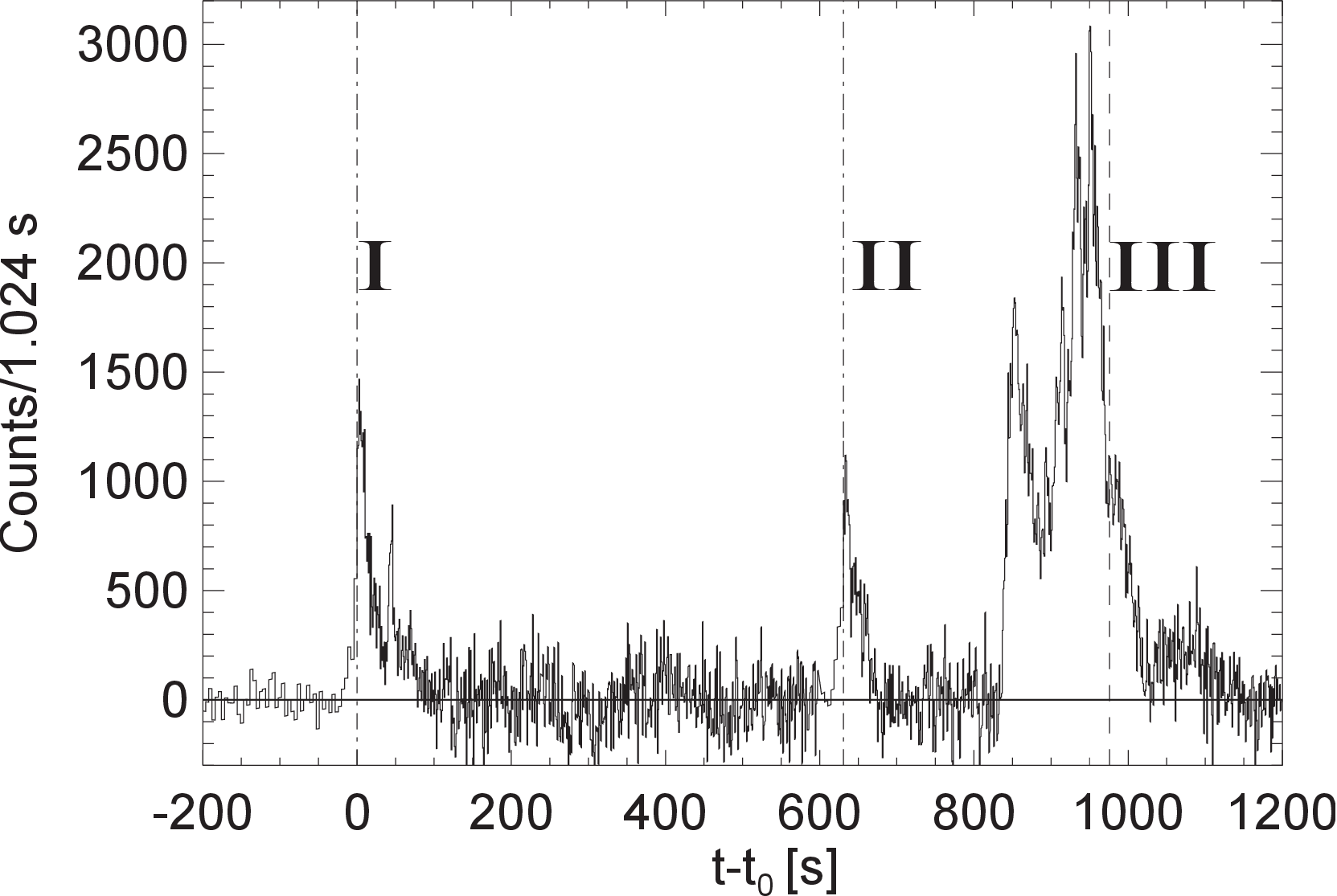}
\caption{Prompt light curve of GRB 091024 background corrected in energy range 
8--1000 keV (Gruber et al., 2011).
}
\label{fig:lcprompt}
\end{figure}

The analysis of the entire prompt emission can be found in Virgili et al. (2013)  and 
Gruber et al. (2011);  here we present only the main results of these analyses 
and we focus on the afterglow emission interpretation.
The background subtracted light curve of prompt emission measured by 
{\it Fermi}/GBM is shown in Fig. \ref{fig:lcprompt}.

There are three emission episodes separated by two periods of quiescence or low activity. 
Because of the irregularity of the background level, from GBM data it is not possible to 
understand if  there is a continuous activity of the central engine between the episodes 
or it is completely switched off. 
Instead, the Konus--Wind data show a low level emission in the period between the pulses
(Virgili et al. 2013).
\vskip 0.2 cm
The main parameters obtained from the analysis of Gruber et al. (2011) and 
Virgili et al. (2013) are shown in Tab. \ref{tab:prompt}. 
These parameters are obtained by fitting the prompt spectrum in each 
episode with a power--law with a high energy cut--off.

\begin{table*}
\centering
\begin{tabular}{|c|c|c|c|c|c|c|}
  \hline
Episode &$t-t_0$   & $T_{90}$ & $E_p$ & $\alpha$ & $\mathcal{F}$ &$E_{\rm ISO}$\\
& [s] & [s] & keV & & [\small$10^{-5} \mbox{ erg}/\mbox{cm}^2]$& [$10^{52}$ erg]\\
  \hline
I & \small $-3.8:67.8$ & \small $72.6 \pm 1.8$ & $412^{+69}_{-53}$ & \small $0.92 \pm 0.07$ & \small $1.81 \pm 0.07$  & $  6.3\pm 0.2$\\
\normalsize II & \small $622.7:664.7$ & \small $44.5 \pm 5.4$ & $371^{+111}_{-71}$ & \small $1.17 \pm 0.07$ & \small $0.79 \pm 0.04$ & $ 2.8\pm 0.1$\\
\normalsize III & \small $838.8:1070.2$ & \small $150 \pm 10$ & $278^{+22}_{-18}$ & \small $1.38 \pm 0.02$ & \small $6.73 \pm 0.09$ &  $ 23.6 \pm 0.3$\\
\hline
\end{tabular}
\caption{Fit parameters of the three episodes of GRB 091024. 
The model fit is a cut--off power--law. Fluence between 8 keV and 40 MeV (Gruber et al., 2011).
}
\label{tab:prompt}
\end{table*} 

In the following section, we interpret the episodes I and II as emission due to the precursors and 
episode III as the main event, using the definition of precursor of Burlon et al. (2008).

\subsection{Optical Afterglow}

\begin{figure*} 
\vskip 0.3 cm
\centering
\includegraphics[scale=1.08, trim=0.7cm 0.9cm 0cm 1cm]{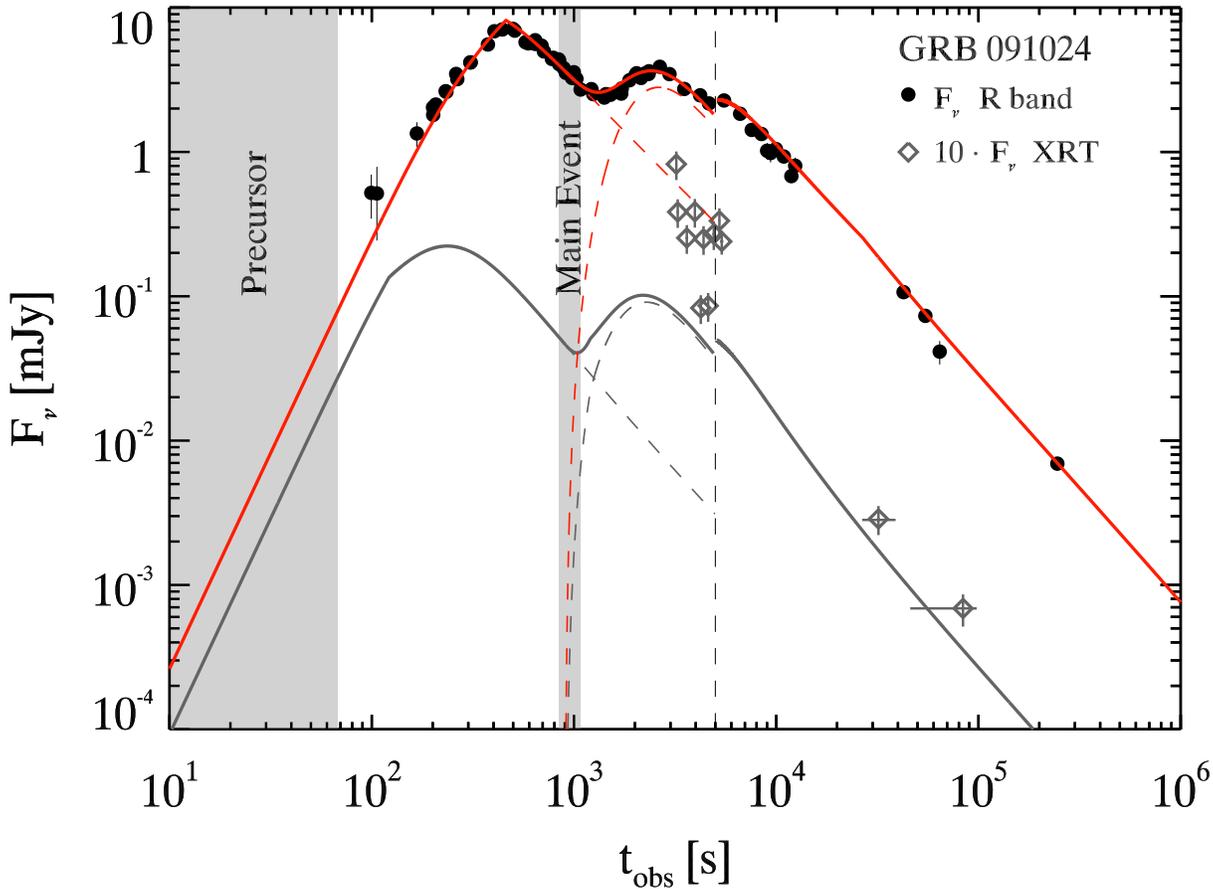}
\vskip 0.5 cm
\caption{
Afterglow light curve of the optical flux in mJy ($R^\prime$ band, $\nu = 4.84 \times 10^{14}$ Hz, black dots) 
of GRB 091024 compared with the proposed model (red solid line). 
The afterglow emission of the precursor and the main event are shown by the red dashed lines.
The vertical black dashed line represents the time when the two fireballs 
(of the precursor and of the main event) collide and merge in a new reborn fireball 
that decelerates in the external CBM.
The grey diamonds show the XRT X--ray emission at 1 keV multiplied by 10, 
compared with the prediction of the model for that frequency.
The grey regions show the time of the precursor and the main event prompt emission.
}
\label{fig:tutto}
\end{figure*}

We study the optical afterglow in $R^\prime$ band (filter peak frequency 
$\nu = 4.84 \times 10^{14}$ Hz).
The light curve is showed in Fig. \ref{fig:tutto}.

There are two main peaks at $t_{\rm obs} \simeq 500$ s and 
$t_{\rm obs} \simeq 2,500$ s, and a small rebrightening at $t_{\rm obs} \simeq 5,000$ s. 

%

\subsection{X--ray Afterglow}

The X--ray emission of GRB 091024 has been observed by XRT only since about $3,000$ s after the trigger. 
The XRT data (flux density at 1 keV) have been retrieved from the {\it Swift} Burst Analyzer. 
X--ray data are shown in Fig. \ref{fig:tutto} (grey diamonds). 
It can be seen that the first part of this light curve is rather different 
from the optical light curve (see also the recent study in D'Avanzo et al. 2012). 
This indicates that the optical and X--ray fluxes have a different origin
(the X--ray flux could be late prompt emission, as suggested by e.g. Ghisellini et al. 2009).
However, the last two X--ray points do describe a decay consistent with the
optical one, indicating the possibility that these two data points belong to the same mechanism
of the optical afterglow (forward external shock).

\subsection{Data Interpretation}

To interpret the optical afterglow data, we applied this procedure dividing the light curve 
in three different time bins.
\begin{enumerate}

\item $t_{\rm obs} \lesssim 900$ s: Here the afterglow light curve is produced only by the 
interaction of the fireball of the 
precursor and the external medium (deceleration of fireball I).

\item $900$ s $ \lesssim t_{\rm obs} \lesssim 5,000$ s: The afterglow light curve is given by the 
superposition of the last part of the afterglow produced by the precursor (fireball I) and the 
one produced by the main event (fireball II).

\item $t_{\rm obs} \gtrsim 5,000$ s: Here the afterglow light curve is produced by the merger 
(fireball I+II) of the two fireballs.
\end{enumerate}
We interpret the first peak as produced by the deceleration of fireball I, 
in the external homogeneous medium. 
Since this optical emission occurs between the prompt emission of the 
precursor and the prompt emission of the main event, we can affirm that it 
\textit{cannot} be the forward shock afterglow emission of the main event, 
since its fireball has not been released yet by the central engine.

The second peak is due to the deceleration of fireball 
II (main event) in the pre--accelerated medium injected by the central engine. 
The third peak corresponds with the merging of fireballs I and II. 
After the merging, the merged fireball I+II continues to decelerate in the external circumburst medium 
(that is not pre--accelerated), so it will produce a more intense and efficient emission. 
This medium jump causes a discontinuity in the light curve, alias the third peak.

The second precursor, formerly labelled as Episode II, is less energetic than the first one, 
and, more importantly, its prompt emission is very close in time to the main event.
We thus expect that its fireball has not enough time to produce its own afterglow before 
being reached by the fireball of the main event.
We confirm this expectation studying the contribution of the fireball of the second precursor in the 
Appendix. There we show that the second precursor afterglow emission can be neglected for small values 
of the initial Lorentz factor of the associated fireball (hereafter labelled as $\Gamma_{0, \rm p,II}$).
Therefore, for the sake of simplicity, we will neglect  the second precursor in the following,
but we add its kinetic energy to the fireball of the main event, i.e.:
\begin{equation}
E_{\rm k, m} = (E_{\rm ISO, II} + E_{\rm ISO, III})/\eta 
\label{eq:ekinet}
\end{equation} 
where $E_{\rm ISO, II}$ and $E_{\rm ISO, III}$ are respectively the isotropic equivalent energy of the Episode II (the second precursor) and the Episode III (the main event) and $\eta$ is the efficiency of conversion of the kinetic energy in radiative energy. The values of $E_{\rm ISO}$ are listed in Tab. \ref{tab:prompt}.

\section{Results}

\subsection{Optical Afterglow results}\label{sec:optres}

Fig. \ref{fig:tutto} shows the optical afterglow light curve compared with the model. 
There is a very good agreement between the model and the data.
The values of the parameters\footnote{Consider that the initial value of the Lorentz bulk 
factor in the third portion (labeled as $\Gamma_{\rm merg}$) is computed with the equation of momentum--energy 
conservation during the merger of the fireballs I and II and it is not a free parameter of the model.}  
used for the solution of the Fig. \ref{fig:tutto} are shown in Tab. \ref{tab:par} 
(fireballs parameters) and in Tab. \ref{tab:mezzo} (medium parameters).
A value $n = 3 \mbox{ cm}^{-3}$ of the circumburst density provides a good agreement between model and data. This value is compatible with the 
known distribution of the circumburst density (see Panaitescu \& Kumar, 2002, Ghisellini et al., 2009 and Cenko et al., 2011).
\begin{table} 
\centering
\begin{tabular}{|c|c|c|c|c|c|}
  \hline
Portion &$\Gamma_0$ &$\Gamma_{\rm merg}$  & $\epsilon_e$ & $\epsilon_B$ & $p$ \\
  \hline
I   & $130$ &...    &  $0.023$ &  $6.0 \times 10^{-4}$  & $2.7$\\
II  & $73$  &...    &  $0.028$ &  $0.017$ &  $2.5$ \\
III &...    &$39.9$ &  $0.0045$ &  $0.003$ &  $2.6$ \\
\hline
\end{tabular}
\caption{Model parameters for the three portions of the light curve: 
I) Precursor afterglow emission, 
II) Main Event afterglow emission, 
III) Afterglow emission produced after the merger of the precursor and the main event fireball.
}
\label{tab:par}
\end{table}

\begin{table} 
\centering
\begin{tabular}{|c|c|c|}
  \hline
$n_0$ & $7.7 \times 10^7$ &cm$^{-3}$ \\
\hline
$s$ & $1$&\\
\hline
$\gamma_0$ & $250$&\\
\hline
$g$ & $0.25$ &\\
\hline
\end{tabular}
\caption{
Parameters of the medium injected by the central engine in the 
quasi--quiescence phases (for $R_0 = 10^{11}$ cm, see Eqs. \ref{eq:profili}).
}
\label{tab:mezzo}
\end{table}

The values of the kinetic energy of the precursor and the main event have been 
computed from the prompt light curve see Tab. 1), assuming an efficiency $\eta = 0.2$,
resulting in $E_{\rm k,p} \simeq 3\times 10^{53}$ erg  and
$E_{\rm k,m} \simeq 1.3\times 10^{54}$ 
erg (according with the Eq. \ref{eq:ekinet}, $E_{\rm k,m}$ includes also the 
energy of the second precursor, previously called as ``Episode II'').

It is not straightforward to explicit analytically the dependence of the predicted flux from the model parameters.
However, we have explored numerically the effects of changing the most relevant parameters 
around the values leading to the curves shown in Fig. \ref{fig:tutto}, listed in
Tabs. \ref{tab:par} and \ref{tab:mezzo}.

The precursor parameters (especially $\Gamma_{0 \rm , p}$) are well constrained by the 
position of the first optical peak (see also \S \ref{sec:timi}, where we show the 
connection between the precursor parameters and the initial bulk Lorentz factor).
Instead, it is more 
challenging to explore the breadth of combinations of the
parameters ($\Gamma_{0, \rm m}$, $\gamma_0$, $g$, $n_0$, $s$)
of the main event, since the corresponding afterglow emission is obtained by the interaction with a non--standard medium. 

\begin{figure} 
\centering
\includegraphics[scale=0.66, trim= 0.9cm 0.5cm 0cm 0cm, angle=0]{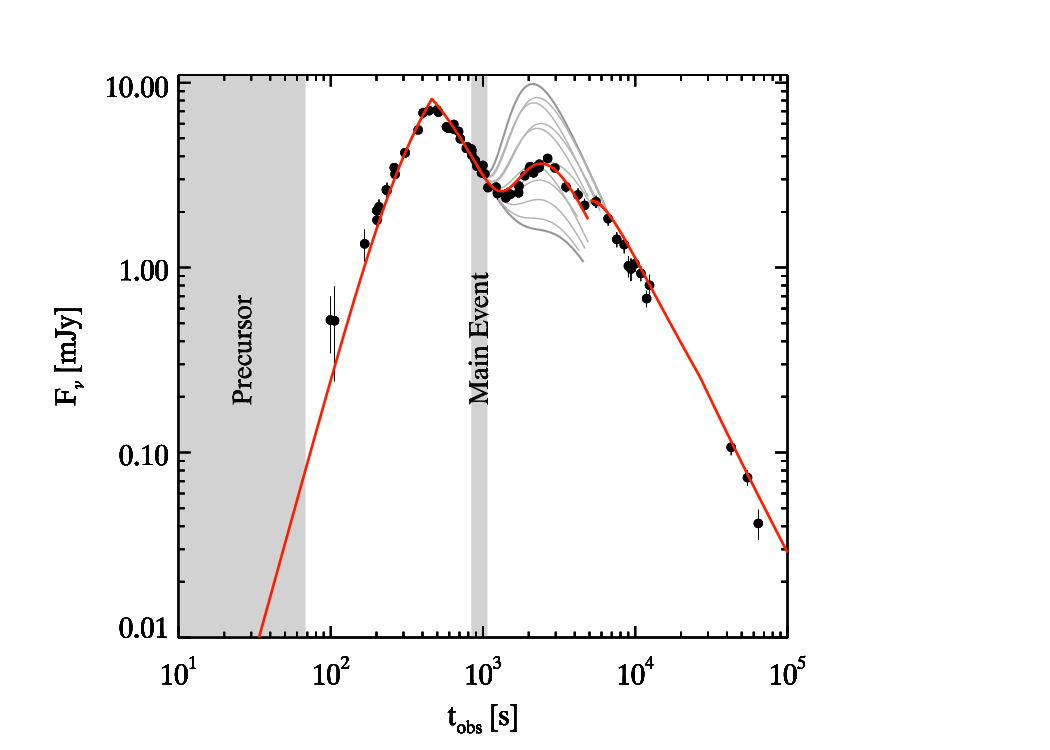}
\caption{
Afterglow light curve of the optical flux
of GRB 091024 compared with the proposed model. The best combination of the parameters is represented 
by the red solid line, the grey lines show how the light curve changes if the parameters are modified in 
a narrow range (see text in \S \ref{sec:optres}). 
The grey regions show the time of the precursor and the main event prompt emission.
}
\label{fig:par}
\end{figure}

To understand how much these parameters are reliable, we examined a narrow range of the space parameter around them,
studying how the second peak of the afterglow changes. 
Fig. \ref{fig:par} shows the variations of the optical light curve if we vary the Lorentz factor parameters 
(i.e. $\Gamma_{0, \rm m}$, $\gamma_0$ and $g$) by  $\sim 4 \%$ and the injected medium density parameters 
(i.e. $n_0$ and $s$) by $\sim 2 \%$ around the best values.
Fig. \ref{fig:par} shows how the light curves are strongly modified even for modest changes of the parameters.
Although we made a good exploration of this particular range of the parameter space, we cannot exclude that with a complete study of the parameter space, we could find other combinations of parameters that can reproduce a similar light curve.

\subsubsection{Precursor analysis}

In this section we will focus on the solution for the optical light curve of the precursor.

\begin{figure} 
\vskip 0.8 cm
\centering
\includegraphics[scale=0.65, trim=0cm 1cm 0cm 0cm]{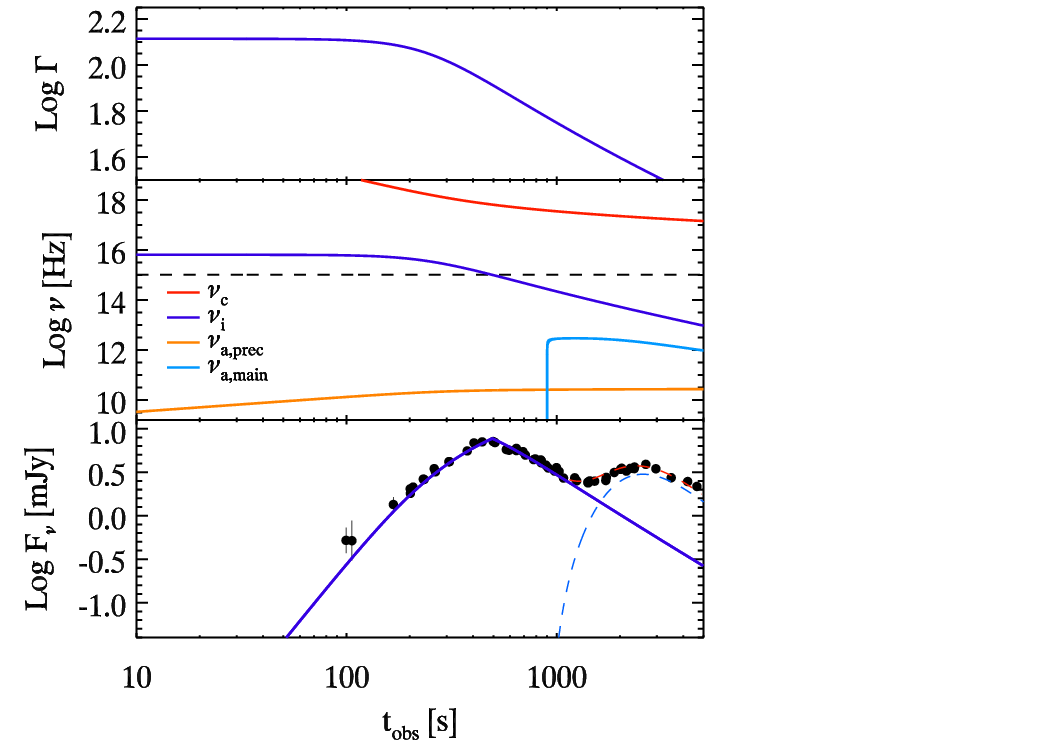}
\vskip 0.3 cm
\caption{
Afterglow emission from the precursor of GRB 091024. 
Top panel: Lorentz factor $\Gamma$ as a function of the observed time $t_{\rm obs}$. 
Mid panel: the injection and cooling rest frame frequencies ($\nu_{\rm i}$ 
and $\nu_{\rm c}$ respectively) 
and the rest frame self--absorption frequencies  
for the precursor shock ($\nu_{\rm a,prec}$) and the main event shock ($\nu_{\rm a,main}$)
as a function of $t_{\rm obs}$. 
The dashed line represents the rest frame frequency of observation 
($\nu = 10^{15}$ Hz). 
All the frequencies are shown rest frame, i.e. corrected for the cosmological redshift.
Bottom panel: optical ($R^\prime$ band) light curve of the afterglow emission of the precursor 
(solid line) and main event light curve (dashed lines). 
Note that the afterglow peak corresponds at the passage of the injection 
frequency $\nu_{\rm i}$.
}
\label{fig:precuro}
\end{figure}

At early times, the flux grows with a slope steeper than $t^2$,
becoming $\propto t^2$ before the peak. 
As described in \S \ref{sec:param}, the fast rising argues in favour of a homogeneous circumburst medium. 
Then there
is a very (and unusual)
sharp peak followed by a flux decay steeper than $t^{-1}$.
In Fig. \ref{fig:precuro} we show a detail of the optical light curve of the precursor 
afterglow, the break frequencies $\nu_{\rm i}$ and $\nu_{\rm c}$ and the Lorentz 
factor of the precursor fireball as a function of the observed time $t_{\rm obs}$.

To reproduce the sharpness of the optical peak, 
we require that it is due not only to the
deceleration of the fireball, 
but
also by 
the transition of the injection frequency $\nu_{\rm i}$.
Moreover, we note that  $\nu_{\rm i} < \nu_{\rm c}$ always, so the afterglow emission of the 
precursor is always in the slow cooling regime.
We stress that the sharpness of the peak,
interpreted as the passage of $\nu_{\rm i}$ simultaneously with
the deceleration time, is not a common property of the afterglow onset,
but a feature of this specific burst.
Therefore, it should not be considered by a weakness of our model.

\subsubsection{Self--absorption}

In \S \ref{sec:break} we claimed that we can neglect the self--absorbed part of the spectrum. 
This approximation is justified only if the observed frequency $\nu$ is larger than the self--absorption
frequency: $\nu > \nu_{\rm a}$.  
We compute $\nu_{\rm a}$ for the precursor and the main event afterglow emission, 
using Eq. \ref{eq:selfabs}. 
The result is shown in the middle panel in Fig. \ref{fig:precuro}.

Both self--absorption frequencies are always much smaller than the frequency of observation $\nu$. 
This justifies our assumed approximation.
Moreover, since the self--absorption frequency of the precursor is smaller than 
the self--absorption frequency
of the second (main event) shock, there is no further absorption
of main event photons by the precursor shock. In conclusion, self--absorption
does not affect the optical emission at all the considered times.

\subsection{A consistency check: the X--ray Afterglow}
We compare the XRT data sampling at 1 keV (Fig. \ref{fig:tutto}, grey diamonds) 
and the light curve at 1 keV  generated by the  same model interpreting
the optical  light curve. 

As mentioned in \S 3.3, at early times the X--ray emission is very different from the optical
counterpart, and is probably due to a different mechanism.
Conversely, the last two points show a decay of the X--ray flux compatible with the 
behaviour of the optical flux and consistent with model predictions. 
Therefore, the model X--ray flux is required to be below the initial X--ray
data points, and be close to the last two X--ray points.
This is indeed what we find, shown in Fig. \ref{fig:tutto}.

\section{Discussion}

Since the general properties of precursors are very similar to the 
ones of the main events, it is natural to assume that precursors and main events
are produced by the same central engine.
The energetics of the precursors is less than, but close to, the 
energetic of the main event, and this implies that the also the precursor
generates its own fireball, that produces the first afterglow.
As far as we know, this is the first time that the consequences of this
scenario are explored.
The onset of the first afterglow is linked to the energetics and the 
bulk Lorentz factor of the precursor, not the ones of the main event, and
to the circumburst density.
The main event launches a second, more powerful, fireball. 
It will run into a {\it moving} medium, producing afterglow emission
not through a standard external forward shock, but with a less efficient
kind of {\it internal} shock.
The onset of this afterglow will then occur at a later time
(the fireball takes longer to decelerate).
The onset time will still depend on the energetics and the bulk Lorentz factor
of the fireball, but in a different way than predicted by the standard
theory.
In addition, it will depend also on the density and velocity of the surrounding medium,
ejected during the quiescent phase (i.e. the time interval between the precursor and the
main event).
Finally, when the second fireball reaches the first, we will have a
third peak of the afterglow light curve.
This can be considered as a refreshed shock.
After the third peak, we have the standard behaviour.

We believe that this three-peak light curve is a distinguishing feature
of the afterglow of the bursts with precursors in this idealized
and simple case.
There could be in fact a variety of cases: 
for instance, the burst could have more than one
precursor (and this would increase the number of peaks in the light curve).
Consider also that the second peak
could not emerge, being hidden by the light curve produced by the first fireball
(especially when the quiescent phase is short), or else the second
fireball might not decelerate before catching the second fireball.

The proposed model well explain the optical afterglow of GRB 091024, which is otherwise 
difficult to understand (see the discussion in Virgili et al. 2013).
Even if admittedly simplified (we have neglected the reverse shock and
treated only the case of an homogeneous CBM), the model accounts for
the observed optical light curve in a quite logical and simple way.

\section*{Appendix: second precursor of GRB 091024}
\begin{figure} 
\centering
\includegraphics[scale=0.66, trim= 0.9cm 0.5cm 0cm 0cm, angle=0]{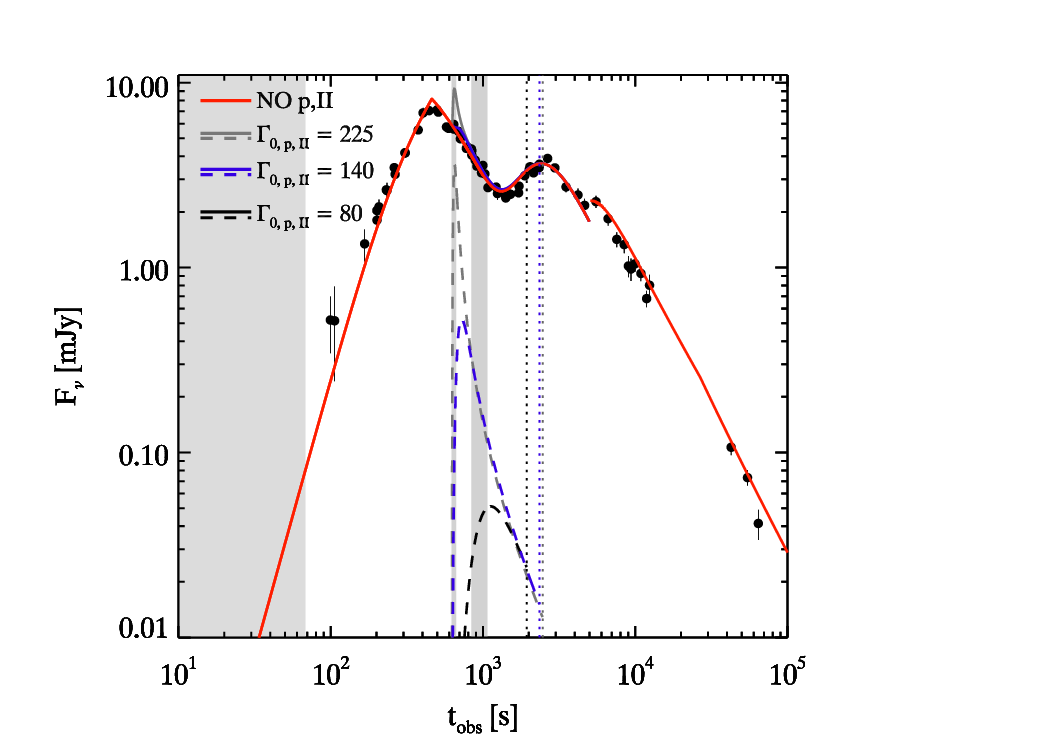}
\caption{
How the afterglow light curve of GRB 091024 changes if we consider also the second precursor. 
Red solid line: the afterglow without the second precursor;  grey dashed line: the 
afterglow of the second precursor with $\Gamma_{0, \rm p,II}=225$,  grey solid line: 
the resulting total afterglow light curve;  blue dashed line: the afterglow of the 
second precursor with $\Gamma_{0, \rm p,II}=140$,  blue solid line: the resulting 
total afterglow light curve;  black dashed line: the afterglow of the second precursor 
with $\Gamma_{0, \rm p,II}=80$,  black solid line: the resulting total afterglow light 
curve (it is almost coincident with the red solid line). 
The vertical dotted lines show when the second precursor fireball is reached by the main event fireball, 
stopping the second precursor afterglow emission. 
They differ according to the $\Gamma_{0, \rm p,II}$ used.
}
\label{fig:prec2}
\end{figure}
In this section we analyse how the optical light curve of GRB 091024 changes 
if we consider also the afterglow produced by the second precursor, formerly labeled as ``Episode II''.

We will show that the contribution of this precursor is negligible as long as its initial bulk 
Lorentz factor ($\Gamma_{0, \rm p,II}$) is smaller than a limit value $\Gamma_{\rm lim}$ that 
is determined numerically.

If we consider also the second precursor in the calculation, the number of parameters increases 
(we must add the initial Lorentz factor of the second precursor fireball 
$\Gamma_{0, \rm p,II}$ and the shock parameters
$\epsilon_{\rm e}$, $\epsilon_{\rm B}$ and $p$), but the number of observable quantities does not change. 
Therefore, the parameters of this second precursor are much less constrained than the other.

We are then forced to use a fixed value for the shock parameters (the same used for the main event 
afterglow emission) and vary the initial Lorentz factor.
Fig. \ref{fig:prec2} shows how the light curve varies if we consider also the emission of 
the second precursor, for different initial Lorentz factors.
We can see that if $\Gamma_{0, \rm p,II} \lsim 140$, the second precursor afterglow is negligible.
In any case, since the fireball of the second precursor is released very close to the fireball 
of the main event, the afterglow of this precursor can not last longer than $\sim 2,300$ s after 
the trigger, and it cannot influence significantly the parameters concerning the
injected medium and the bulk Lorentz factor of the fireball associated to the main event.

\section*{Acknowledgments}
We thank the anonymous referee for constructive comments.
We thank D. Gruber for discussion. PRIN--INAF 2011 is acknowledged for financial support.

\end{document}